\documentclass[manuscript]{aastex63}

\newcommand\solarmass{$M_\odot$}
\newcommand\vkm{km\,s$^{-1}$}

\shorttitle{Feedback from $\gamma$~Cas}
\shortauthors{Chen et al.}
\begin{document}

\title{Feedback from $\gamma$~Cassiopeiae: Large Expanding Cavity, Accelerating Cometary Globules, and Peculiar X-ray Emission}

\correspondingauthor{Xuepeng Chen}
\email{xpchen@pmo.ac.cn}

\author{Xuepeng~Chen}
\author{Weihua~Guo}
\author{Li~Sun}
\author{Jiangchen~Feng}
\affiliation{Purple Mountain Observatory \& Key Laboratory of Radio Astronomy, Chinese Academy of Sciences, 10 Yuanhua Road, 210023 Nanjing, China}
\affiliation{School of Astronomy and Space Science, University of Science and Technology of China, Hefei, Anhui 230026, China}

\author{Yang~Su}
\author{Yan~Sun}
\author{Shaobo~Zhang}
\author{Xin~Zhou}
\author{Qing-Zeng~Yan}
\affiliation{Purple Mountain Observatory \& Key Laboratory of Radio Astronomy, Chinese Academy of Sciences, 10 Yuanhua Road, 210023 Nanjing, China}

\author{Min~Fang}
\author{Ji~Yang}
\affiliation{Purple Mountain Observatory \& Key Laboratory of Radio Astronomy, Chinese Academy of Sciences, 10 Yuanhua Road, 210023 Nanjing, China}
\affiliation{School of Astronomy and Space Science, University of Science and Technology of China, Hefei, Anhui 230026, China}

\begin{abstract}

We present wide-field multi-wavelength observations toward $\gamma$ Cassiopeiae (or $\gamma$~Cas for short) in order to study its feedback toward interstellar
environment. A large expanding cavity is discovered toward $\gamma$~Cas in the neutral hydrogen (HI) images at a systemic velocity of about --10\,\vkm. The 
measured dimension of the cavity is roughly 2\fdg0\,$\times$\,1\fdg4 (or 6.0\,pc\,$\times$\,4.2\,pc at a distance of 168\,pc), while the expansion velocity is 
$\sim$\,5.0\,$\pm$\,0.5\,\vkm. The CO observations reveal systematic velocity gradients in IC\,63 ($\sim$\,20\,\vkm\,pc$^{-1}$) and IC\,59 ($\sim$\,30\,\vkm\,pc$^{-1}$), 
two cometary globules illuminated by $\gamma$~Cas, proving fast acceleration of the globules under stellar radiation pressure. The gas kinematics indicate that 
the cavity is opened by strong stellar wind, which has high potential to lead to the peculiar X-ray emission observed in $\gamma$~Cas. Our result favors a recent 
new scenario that emphasizes the roles of stellar wind and binarity in the X-ray emission of the $\gamma$~Cas stars.

\end{abstract}


\section{Introduction} 

$\gamma$~Cas is the prototype of Be stars (Secchi 1866; Rivinius et al. 2013) and is also one of the most studied stars in the sky (see, e.g., Poeckert \& 
Marlborough 1978; Henrichs et al. 1983; Stee et al. 1995; Smith \& Robinson 1999). It is a bright B0.5\,IVe star located at a distance of 168\,$\pm$\,4\,pc 
from the Sun (van~Leeuwen 2007). Radial velocity (RV) observations indicate that $\gamma$~Cas is a binary system with an orbital period of  
$\sim$\,203.5\,days (Harmanec et al. 2000; Nemravov{\'a} et al. 2012). The primary stellar mass is estimated to be $\sim$\,15\,\solarmass\ from its effective 
temperature (Harmanec et al. 2000), while the companion is suggested to be a white dwarf  (Harmanec et al. 2000) or a helium star (Nemravov{\'a} et 
al. 2012), with an estimated mass of about 1.0\,\solarmass.

It has been known that massive stars strongly influence their surrounding environment throughout their lifetime via strong UV radiation, stellar wind, and 
eventually a supernova explosion (see, e.g., Krumholz et al. 2014; Dale 2015). Previous observations toward the interstellar environment of $\gamma$~Cas 
focused on IC\,63 and IC\,59, two cometary globules brightened by $\gamma$~Cas (e.g., Blouin et al. 1997; Karr et al. 2005), which are located to the 
northeast (IC\,63) and north (IC\,59) of $\gamma$~Cas, respectively. Both IC63 and IC59 are also associated with S185 (see Sharpless 1959), which 
is a bright H\,II region with a shell to the northeast of $\gamma$~Cas (see, e.g., Sun et al. 2007). The two globules were well-studied, in order to understand 
the physical and chemical effects from stellar UV radiation (see, e.g., Jansen et al. 1994; Andersson et al. 2013; Andrews et al. 2018), as well as the 
magnetic field structure due to radiative grain alignment (see Soam et al. 2017, 2021).

In this work, we present wide-field multi-wavelength images toward $\gamma$~Cas, based on the observational data obtained from the neutral hydrogen 
(HI) 4$\pi$ survey (HI4PI; see HI4PI Collaboration 2016) and the CO Galactic plane survey conducted with the Purple Mountain Observatory (PMO) 13.7\,m 
millimeter-wavelength telescope. Complementary images from the full-sky H$\alpha$ map (Finkbeiner 2003) and the all-sky survey by the Wide-field Infrared 
Survey Explorer (WISE; Wright et al. 2010) are also used for comparisons with the HI and CO line data.

\section{Observations and data reduction} 

\subsection{HI4PI HI data}

We retrieved the HI line data from the HI4PI survey for the $\gamma$~Cas region. The HI4PI survey is based on data from the recently completed first 
coverage of the Effelsberg-Bonn HI Survey (EBHIS; Winkel et al. 2016) and from the third revision of the Galactic All-Sky Survey (GASS; Kalberla \& Haud 
2015). The HI4PI survey toward the northern sky was performed with the Effelsberg 100\,m telescope in Germany and the southern sky performed with the 
64\,m Parkes radio telescope in Australia. The GASS and EBHIS data were reduced by the two survey groups independently (see, e.g., Kalberla \& 
Haud 2015 for GASS and Winkel et al. 2016 for EBHIS, respectively). After detailed comparisons and analyses, the two data sets were merged together 
with the same angular and spectral resolutions. The released HI4PI data have a velocity resolution of $\sim$\,1.29\,\vkm\ and an angular resolution of 
$\sim$\,16\farcm2, while the typical rms noise of the line data is about 43\,mK. For more details, we refer to the HI4PI survey introduction paper (see 
HI4PI Collaboration 2016). Figure~5 in Appendix~A shows the HI4PI HI line velocity channel maps toward $\gamma$~Cas.

\subsection{CO Observations and Data Reduction}

The CO\,(1--0) line observations toward the $\gamma$~Cas region is part of the Milky Way Imaging Scroll Painting (MWISP) project for 
investigating the nature of the molecular gas along the northern Galactic plane (with Galactic latitude $|b|$\,$\leq$\,5$^\circ$), using the PMO 
13.7\,m millimeter-wavelength telescope at the Delingha station in Qinghai, China (see, e.g., Su et al. 2019). The CO observations were 
made from 2011 November to 2020 February. The nine-beam Superconducting Spectroscopic Array Receiver (SSAR; Shan et al. 2012) worked 
as the front end in sideband separation mode. Three CO\,(1--0) lines were simultaneously observed, $^{12}$CO at the upper sideband (USB) and 
two other lines, $^{13}$CO and C$^{18}$O, at the lower sideband (LSB). The total of the pointing and tracking errors is about 5$''$, and the 
half-power beam width (HPBW) is $\sim$\,55$''$ ($^{12}$CO line). A fast Fourier transform (FFT) spectrometer with a total bandwidth of 1000\,MHz 
and 16,384 channels was used as the back end. The corresponding velocity resolutions were $\sim$\,0.16\,\vkm\ for the $^{12}$CO line and 
$\sim$\,0.17\,\vkm\ for both the $^{13}$CO and C$^{18}$O lines. More details about the PMO 13.7\,m telescope system are described in the 
telescope status reports\footnote{See http://www.radioast.csdb.cn/zhuangtaibaogao.php}.

After removing bad channels and abnormal spectra, and correcting the first-order (linear) baseline fitting, the data were re-gridded into standard 
FITS files with a pixel size of 30$\arcsec$\,$\times$\,30$\arcsec$ (approximately half of the beam size). The average rms noises of all final spectra 
are about 0.5\,K for $^{12}$CO  and about 0.3\,K for $^{13}$CO and C$^{18}$O. Finally, we mosaicked the data cubes toward the $\gamma$~Cas 
region to analyze the morphology and physical properties of molecular gas. Figures~6 and 7 in Appendix show the $^{12}$CO line velocity channel 
maps toward $\gamma$~Cas.

\section{Results and Discussion} 

\subsection{A Large Expanding Cavity toward $\gamma$~Cas}

\begin{figure*}
\begin{center}
\includegraphics[width=17.0cm,angle=0]{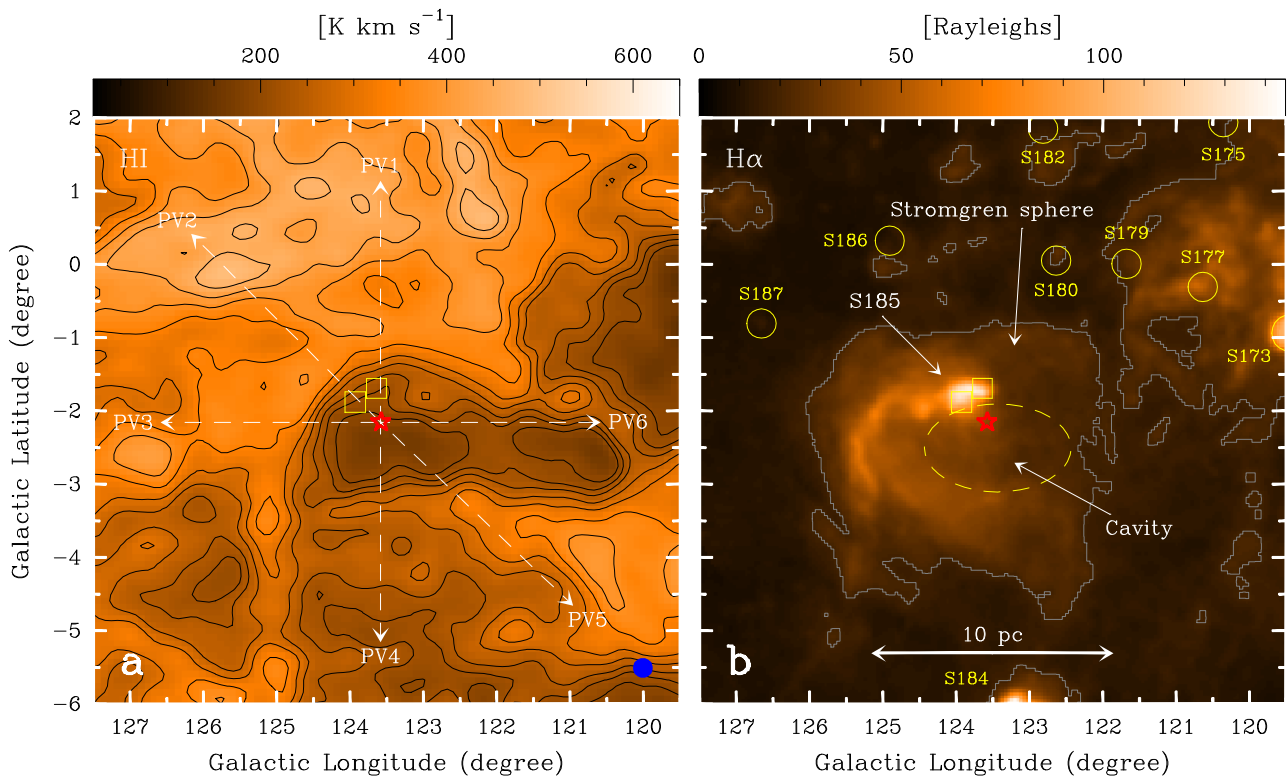}
\end{center}
\caption{\footnotesize (a) Wide-field HI4PI HI intensity images toward $\gamma$~Cas, integrated between --13.0 and --7.0\,\vkm. The contours represent 
the emission at 200, 220, 240, 270, and 300\,K\,\vkm, and then increase by steps of 40\,K\,\vkm. The red stellar symbol marks the position of $\gamma$~Cas. 
White dashed arrow lines show cutting routings to extract the position-velocity diagrams. Yellow squares show the positions of IC\,63 and IC\,59 globules,
respectively. Blue beam shows the angular resolution (16\farcm2) of the HI4PI data. (b) The H$\alpha$ image toward $\gamma$~Cas. The yellow dashed ellipse 
shows the cavity found in the HI images. The H$\alpha$ contour (grey) represents the emission at 18 Rayleighs (Rayleigh = 10$^6$/4pi photons/cm$^2$/s/sr), 
outlining a large Str{\"o}mgren sphere with a radius of $\sim$\,1\fdg8-2\fdg0. Yellow circles mark the positions of other H\,II regions in the field. These HII 
regions, such as S173 ($\sim$\,--35\,\vkm), S177 ($\sim$\,--34\,\vkm), and S187 ($\sim$\,--15\,\vkm), have different systemic velocities (see Blitz et al. 1982) and 
are located much farther away than S185 (associated with $\gamma$~Cas).
\label{HI4PI}}
\end{figure*}

Long-term (over 6000 days or 16.84 yr) RV monitors of $\gamma$~Cas showed a spread of velocities in an approximate range of --17 to --1\,\vkm\ in 
the local standard of rest (LSR) frame (see Nemravov{\'a} et al. 2012).  Taking an average of the RV measurements, the systemic velocity of $\gamma$~Cas 
was estimated to be --10\,\vkm, which is consistent with the systemic velocities derived after the epoch of HJD 245 2000 (Nemravov{\'a} et al. 2012; see also
Pollmann 2016). After checking the HI4PI line maps toward $\gamma$~Cas (see HI velocity channel maps in Appendix~A), we find a cavity within the 
measured RV range of $\gamma$~Cas, which is most prominent at its systemic velocity (i.e., --10\,\vkm). Figure~1a shows the HI intensity image toward 
$\gamma$~Cas, integrated between --13.0 and --7.0\,\vkm. The cavity is extended in the east-west direction. We note that no other massive OB star is 
found within the cavity region in the Galactic local arm, except $\gamma$~Cas\footnote{After checking the SIMBAD Astronomical Database (see 
http://simbad.u-strasbg.fr/simbad/), another Be star, HD 4931 ($l$\,=\,123\fdg034, $b$\,=\,--2\fdg782), is seen toward the HI cavity. Its radial velocity is 
$\sim$\,52.2\,\vkm, and observed parallax is 0.5541 mas (or a distance of 1800 pc) by the {\it Gaia} satellite.}.

\begin{figure*}
\begin{center}
\includegraphics[width=17.0cm,angle=0]{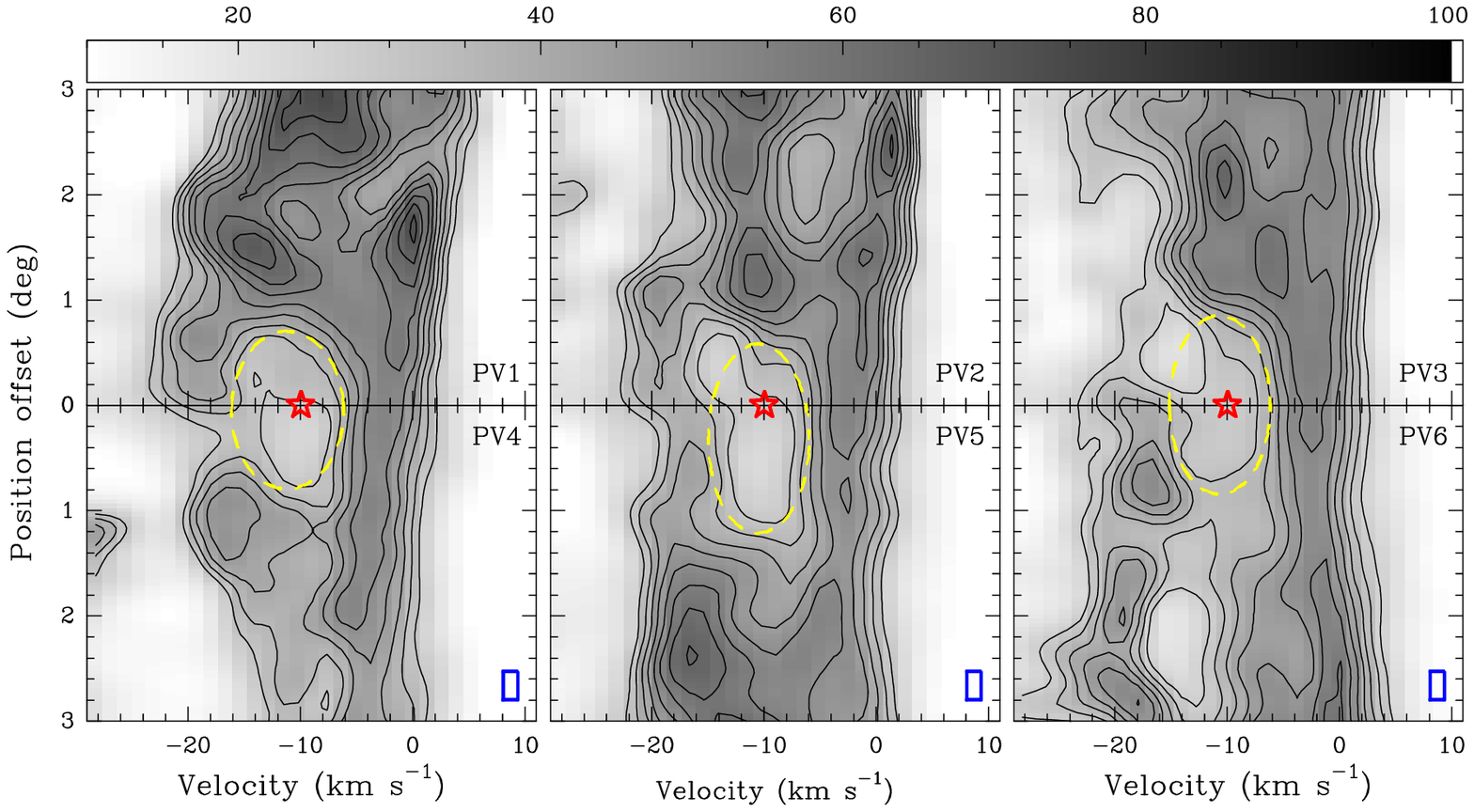}
\end{center}
\caption{\footnotesize The HI position-velocity diagrams cutting across the cavity toward $\gamma$~Cas. In all the panels, red stellar symbol marks the 
position of $\gamma$~Cas. Blue rectangle shows the angular resolution (16\farcm2) and velocity resolution (1.29\,\vkm) in the HI4PI observations. The HI
contours start at 30\,K and then increase in steps of 5\,K (where 1$\sigma$\,$\sim$\,0.2\,K). The yellow dashed ellipses show fitting results toward the 
cavity-like patterns seen in the diagrams.\label{HI4PI-PV}}
\end{figure*}

Figure~1b shows wide-field H$\alpha$ image toward $\gamma$~Cas. In the H$\alpha$ image, extended ionized hydrogen emission nebula is seen around 
$\gamma$~Cas, with a bight shell toward the east and northeast of $\gamma$~Cas. This bright shell was also seen in the $\lambda$\,6\,cm radio continuum 
observations (Sun et al. 2007). The large extended H$\alpha$ emission nebula, with a radius of  $\sim$\,1\fdg8-2\fdg0, was suggested to represent the 
Str{\"o}mgren sphere produced by the photoionization from $\gamma$~Cas (see Karr et al. 2005 and Appendix~B).
It is widely known that massive stars experience strong mass loss in the form of winds, which will sweep up the interstellar medium (ISM) and create low 
density wind-blown cavities surrounded by dense shells (see, e.g., Weaver et al. 1977; Freyer et al. 2003). The comparison between the HI and H$\alpha$ 
images shows that the HI cavity discovered toward $\gamma$~Cas is located within the large H$\alpha$ nebula caused by strong photoionization, which is 
in good agreement with the classical picture for the feedback from massive stars\footnote{In the classical picture, the radiation field of a massive star first 
dissociates the interstellar gas and forms a photodissociation region of neutral hydrogen. Subsequently, the Lyman continuum photons of the star ionize the 
HI gas and produce an H\,II region that expands into the neutral ambient medium. Then, a fast stellar wind creates shocks that form a wind-blown cavity filled 
with hot plasma, which expands into the H\,II region (see, e.g., Weaver et al. 1977 and Freyer et al. 2003 for more details).}. Therefore, we suggest that 
the HI cavity is produced by stellar wind from $\gamma$~Cas.

Figure~2 shows the HI position-velocity (PV) diagrams along the routings across the cavity toward $\gamma$~Cas (see the routings shown in Fig.\,1a). 
Cavity-like PV patterns are detected in the various directions, indicating that the cavity is expanding. As seen in the diagrams, the cavity-like PV patterns 
could be well fitted by ellipses, with the position and velocity radii representing the projected radii and expanding velocities of the cavities, respectively. 
The measured expanding velocity of the cavity is roughly 5.0\,$\pm$\,0.5\,\vkm, while the radii of the cavity, depending on the azimuths, range from 
$\sim$\,0\fdg7 (or $\sim$\,2.1\,pc at distance of 168\,pc) to $\sim$\,1\fdg0 ($\sim$\,3.0\,pc). Using the equation (3) in Churchwell et al. (2006), 
the eccentricity of the cavity is calculated to be roughly 0.71, which is a typical value in the Galactic cavity/bubble statistics (see, e.g., Churchwell et al. 2007). 
This kind of extended structure is generally explained by non-uniform ambient ISM into which the cavities are expanding and/or anisotropic stellar wind and 
radiation field\footnote{To the west of the $\gamma$~Cas HI cavity is another cavity-like structure in the HI images (see Figure 1a). This cavity may be 
opened by other bright stars in the region. In this work we focus on the HI cavity toward $\gamma$~Cas.}.

Adopting the standard method (Weaver et al. 1977), the value of the mechanical luminosity (the kinetic energy of stellar wind per unit time) of the wind 
($L_{\rm wind}$) can be calculated by 
\begin{equation}\label{f.luminosity}
 L_{\rm wind}~\approx ~\frac{1}{3}(\frac{{\it n_{\rm gas}}}{\rm cm^{-3}})~(\frac{{\it R_{\rm c}}}{\rm pc})^{2}~(\frac{{\it V_{\rm c}}}{\rm km\,s^{-1}})^{3} \times\,10^{30}~{\rm erg\,s^{-1}},
\end{equation}
in order to excavate a cavity with a radius of $R_{\rm c}$ and an expansion velocity of $V_{\rm c}$ within a cloud with a density of $n_{\rm gas}$.
The density $n_{\rm gas}$ can be measured from the radius and column density of the cavity by
\begin{equation}\label{f.density}
 n_{\rm gas}~ = ~\frac{3{\it N_{\rm shell}}}{{\it R_{\rm c}}},
\end{equation}
where $N_{\rm shell}$ is the column density observed at the shell of the cavity. Based on the HI line observations, the density $n_{\rm gas}$ of the hydrogen 
gas surrounding $\gamma$~Cas is measured to be $\sim$\,150\,cm$^{-3}$. With the radii and expansion velocity measured above, the $L_{\rm wind}$ 
is then estimated to be $\sim$\,(3.9\,$\pm$\,1.0)\,$\times$\,10$^{34}$\,erg\,s$^{-1}$ for the $\gamma$~Cas cavity.
By the relation of $t_{\rm kin}$ = $\frac{16}{27}$$\frac{R_{\rm c}}{\rm pc}$$\frac{\rm km\,s^{-1}}{V_{\rm c}}$\,$\times$\,10$^6$\,yr (see Weaver et al. 1977),
the kinetic timescale $t_{\rm kin}$ of the wind needed for opening such a cavity  is estimated to be $\sim$\,(3.0\,$\pm$\,0.5)\,$\times$\,10$^5$\,yr. 

The mechanical luminosity of the wind is also defined by $L_{\rm wind}$~=~$\frac{1}{2}$$\dot{M}$$V$$^2_{\rm wind}$, where $\dot{M}$ and $V_{\rm wind}$ 
is the mass-loss rate and velocity of the wind, respectively, which can be measured through optical and/or infrared spectroscopies. For $\gamma$~Cas, 
the measured stellar wind is strong, with a fast velocity of $\sim$\,1500-1800\,\vkm\ (Smith \& Robinson 1999) and a mass-loss rate of 
5\,$\times$\,10$^{-8}$\,$M_{\odot}$\,yr$^{-1}$ (Stee et al. 1995), and the calculated $L_{\rm wind}$ is $\sim$3.6-5.1\,$\times$\,10$^{34}$\,erg\,s$^{-1}$. 
Therefore, the mechanical luminosities of the wind observed from cavity gas kinematics and stellar spectroscopies are consistent with each other.

\subsection{The Cometary Globules in Acceleration}

\begin{figure*}
\gridline{\fig{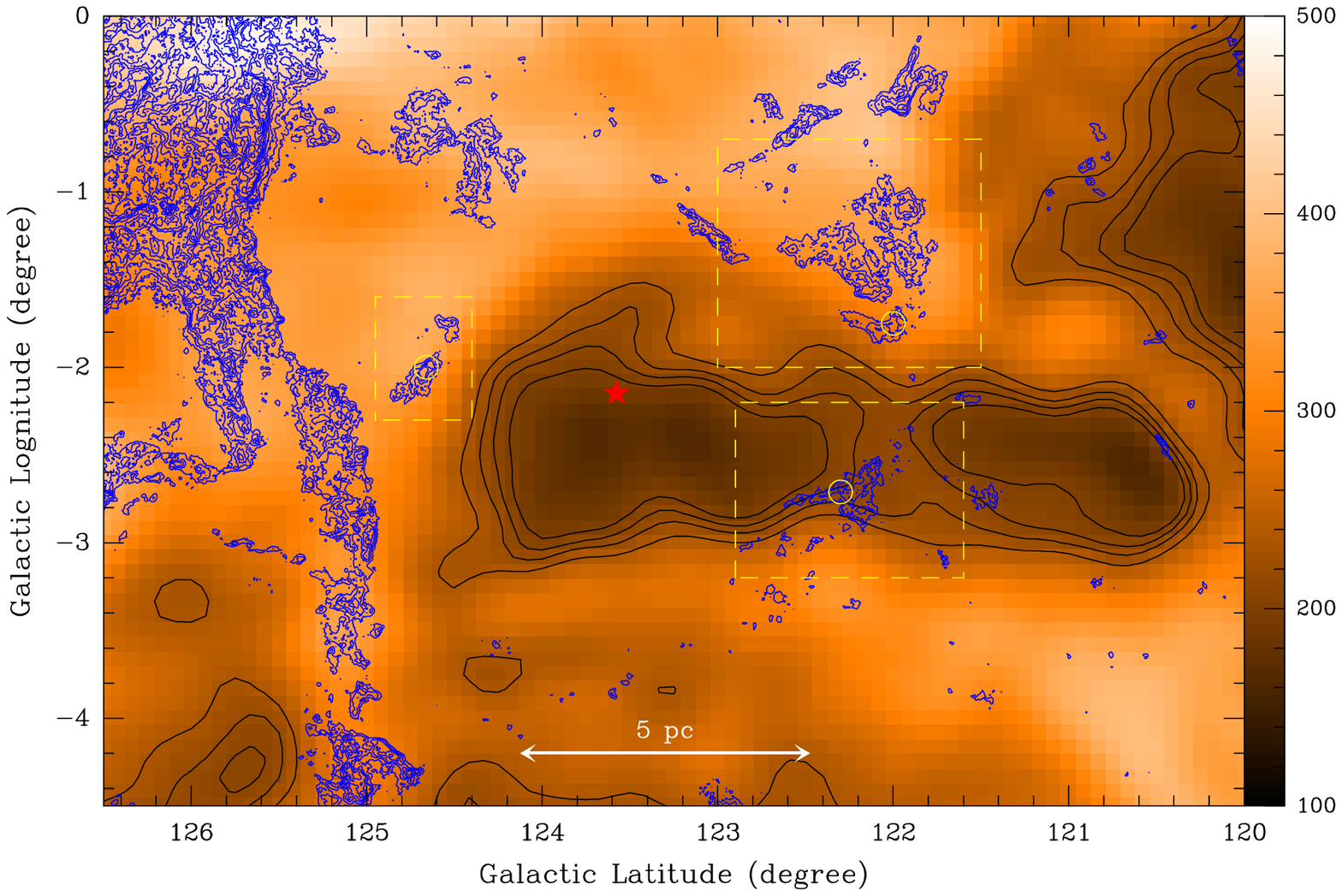}{0.6\textwidth}{(a)}}
\gridline{\fig{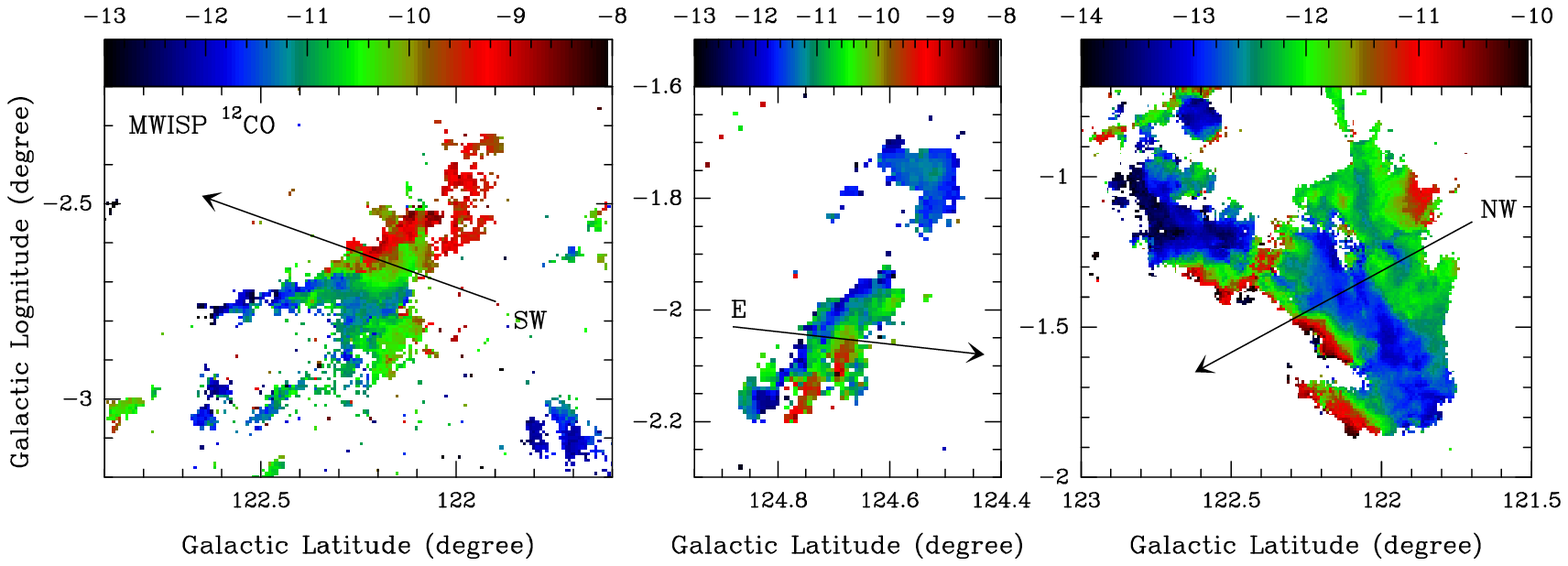}{0.75\textwidth}{(b)}}
\gridline{\fig{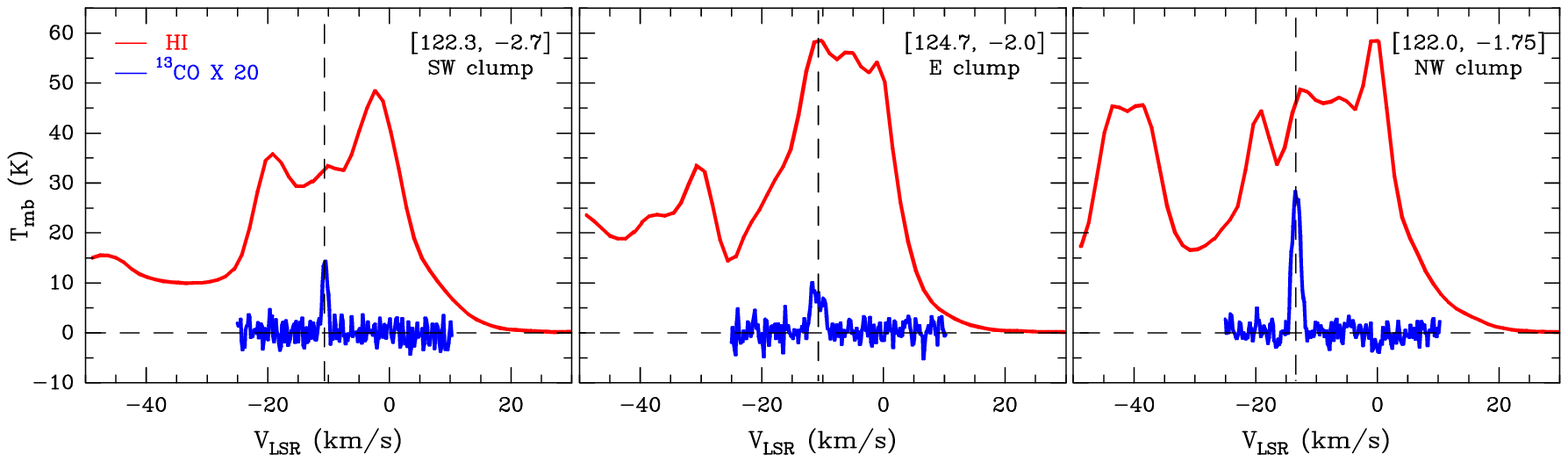}{0.75\textwidth}{(c)}}
\caption{\footnotesize (a) The HI4PI HI intensity image, overlapped with the MWISP $^{12}$CO contours (blue). The integrated velocity range
is [--13, --7] for both the HI and CO lines. The HI contours (black) represent the emission at 200, 210, 220, and 235\,K\,\vkm, respectively. The
$^{12}$CO contours start from 5\,$\sigma$ level and then increase by steps of 5\,$\sigma$ (1\,$\sigma$\,$\sim$\,0.6\,K\,\vkm). The yellow dashed
 rectangles show the extents of the CO clumps around the cavity. (b) The MWISP $^{12}$CO velocity fields of the clumps. The unit of the scale 
 bar is \vkm. The arrow line shows the direction toward $\gamma$~Cas for individual clumps. (c) The HI4PI HI and MWISP $^{13}$CO spectra 
 of the clumps, sampled at the positions marked by yellow circles in panel a.
\label{fig:HI-CO}}
\end{figure*}

\begin{figure*}
\begin{center}
\includegraphics[width=12.0cm,angle=0]{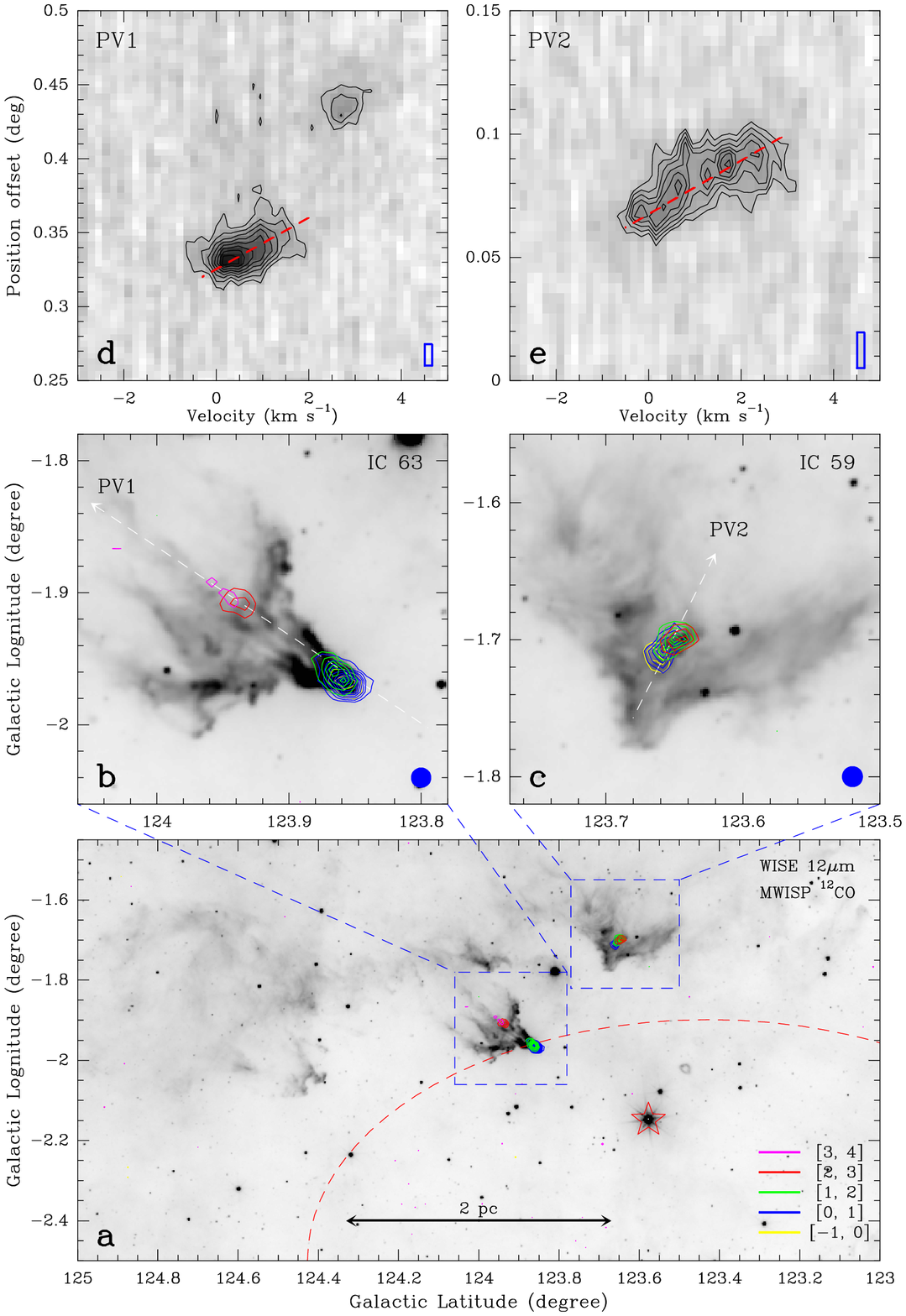}
\end{center}
\caption{\scriptsize (a) The wide-field 12$\mu$m image, overlapped with the MWISP $^{12}$CO contours. The CO contours with different colors represent 
the emission integrated within different velocity ranges (see the image). All contours start from 4\,$\sigma$ level and increase by a step of 2\,$\sigma$ level 
($\sigma$ $\sim$\,0.25\,K\,\vkm). The stellar symbol marks the position of $\gamma$~Cas. The red dashed ellipse shows the cavity found in the HI images. 
(b) The enlarged view for the IC\,63 globule. All contours start from 4\,$\sigma$ and increase by a step of 3\,$\sigma$. The white dashed line shows the cutting 
routing to extract the PV diagram across the globule. The blue beam shows the angular resolution (55$''$) in the CO observations. (c) Similar to panel b, but 
for the IC\,59 globule, and CO contours increasing step is 2\,$\sigma$. (d) The PV diagram along the routing shown in panel b. The contours start from 3\,$\sigma$ 
and increase by a step of 2\,$\sigma$, where $\sigma$ = 0.2\,K. The blue rectangle indicates the angular resolution (55$''$) and velocity resolution (0.16\,\vkm) 
in the $^{12}$CO observations. (e) The PV diagram along the routing shown in panel c. The contours start  from 3\,$\sigma$ and increase by a step of 
1\,$\sigma$.\label{MWISP}}
\end{figure*}

Both observational and numerical studies suggested that molecular gas could be formed in dense shells of large cavities/bubbles created by massive stars
(see, e.g., Dawson et al. 2013, Inutsuka et al. 2015, and references therein). Numerical studies suggest that the timescale for accumulating CO molecular 
cloud ($t_{\rm acc}$) from HI atomic gas behind shock waves can be estimated by {\it nv}~$\geq$~20(20\,Myr/$t_{\rm acc}$)\,cm$^{-3}$\,\vkm, where $n$ is the 
HI gas density and $v$ is the velocity of the shock wave (see Bergin et al. 2004). Given $n$\,=\,150\,cm$^{-3}$ (see above) and assuming a typical sound 
speed of 10\,\vkm, the estimated $t_{\rm acc}$ is roughly 1.3--2.7\,$\times$\,10$^5$\,yr (when shock velocity $v$ is in the range of 10--20\,\vkm), which is 
comparable to the the kinetic age of the $\gamma$~Cas cavity ($t_{\rm kin}$\,$\sim$\,3.0\,$\pm$\,0.5\,$\times$\,10$^5$\,yr; see above). 

Figure~3a shows the wide-field MWISP CO intensity image toward $\gamma$~Cas, integrated between --13.0 and --7.0\,\vkm\ (see also the CO velocity channel 
map in Appendix~A). A few CO clumps are detected in the dense shell of the HI cavity toward $\gamma$~Cas (see Figure~3a). After inspecting the CO velocity 
fields, we find in these clumps systematic velocity gradients toward $\gamma$~Cas (see Figure~3b), and measured velocity gradients in these clumps range from 
$\sim$\,4 to $\sim$\,7\,\vkm\,pc$^{-1}$. This suggests that these clumps are affected by the stellar wind from $\gamma$~Cas. We further check the HI-CO spectra 
from these clumps to search for the signature of HI narrow self-absorption (HINSA). HINSA is generally accepted as the evidence for the presence of cold HI gas, 
which suggests the formation phase of molecular gas, when HINSA is spectrally and spatially coincident with molecular line emission (see, e.g., Li \& Goldsmith 
2003; Goldsmith \& Li 2005). Nevertheless, no clear HINSA signature is found in these clumps (see Figure~3c). Higher angular-resolution and sensitivity HI and 
CO line observations are needed to further study whether there is molecular gas formation in the shell of the $\gamma$~Cas cavity.

Figure~4a shows the MWISP $^{12}$CO line emission images (contours) toward IC\,63 and IC\,59, plotted on the WISE 12\,$\mu$m image. As seen in the figure, 
strong CO gas emission is detected from both the IC\,63 and IC\,59 globules, at the velocity range between $-$1\,\vkm\ and 4\,\vkm\  (see also the CO velocity
channel map in Appendix~A). It is of interest to note that the peak CO emission velocities of IC\,63 ($\sim$\,0.5\,\vkm) and IC\,59 ($\sim$\,1.0\,\vkm) are $\sim$\,10\,\vkm\ 
offset from the systemic velocity of $\gamma$~Cas (--10\,\vkm). Nevertheless, the velocity offsets are actually expected, as cometary globules should be accelerated 
by the radiation of its exciting star (see, e.g., Lefloch \& Lazareff 1994). For $\gamma$~Cas (a bright B0.5\,IVe star), the estimated velocity difference due to the 
radiation acceleration is approximately 10\,\vkm\ between cometary globules and the exciting source (Soam et al. 2017), which is the exact velocity offset seen in the 
CO and HI observations toward $\gamma$~Cas. 

In the MWISP $^{12}$CO observations, clear and systematic velocity gradients are revealed along the IC\,63 ($\sim$\,20\,\vkm\,pc$^{-1}$) and IC\,59 ($\sim$\,30\,\vkm\,pc$^{-1}$) 
globules, proving the acceleration of the two globules. The observed velocity gradients in the two globules are larger than those measured in the CO clumps around the cavity 
(see above), and also much larger than those observed in other cometary globules, such as CG\,7S ($\sim$\,3\,\vkm\,pc$^{-1}$; Lefloch \& Lazareff 1995), IC\,1369 
($\sim$\,5.5\,\vkm\,pc$^{-1}$; Sugitani et al. 1997), and the  Eagle Nebula's fingers ($\sim$\,1.7\,\vkm\,pc$^{-1}$; White et al. 1999). In this context, we consider that 
the IC\,63 and IC\,59 cometary globules, under strong radiation pressure from $\gamma$~Cas, were pre-existing before the HI cavity.

\subsection{The Peculiar X-ray Emission from $\gamma$~Cas}

$\gamma$~Cas was also one of the first extra solar X-ray sources discovered (Jernigan et al. 1976; Mason et al. 1976). However, its  X-ray emission is 
significantly different from that observed from the bulk of Be stars (see, e.g., a review by Smith et al. 2016 and references therein). Galactic X-ray surveys 
have found a growing number of Be stars with similar X-ray properties and sharing a narrow range of characteristics (Smith et al. 2016; Naz{\'e} et al. 
2020), such as hard X-ray emission, multiple thermal components, high variability, and moderated X-ray luminosity ($\sim$\,10$^{31}$--10$^{32}$\,erg\,s$^{-1}$). 
The origin of the peculiar X-ray emission of the $\gamma$~Cas stars is one of the puzzles in X-ray stellar astrophysics, which is important 
to understand massive star evolution and the energy budget of the Galaxy. Two competing scenarios, magnetic star-disk interaction\footnote{In this 
scenario, magnetic field anchored in the rotating Be star interacts with its Keplerian circumstellar disk. Reconnection events accelerate particles that
 produce X-rays when impacting the disk or the photosphere.} (e.g., Motch et al. 2015) and accretion onto compact companion\footnote{In binaries, 
 accretion of matter onto a compact object, likely a white dwarf, could power strong X-ray luminosity.} (e.g., Hamaguchi et al. 2016), have been suggested 
 to explain these outstanding features, but both providing imperfect matches with observations (Naz{\'e} et al. 2017, 2019; Langer et al. 2020).

Considering that $\gamma$~Cas stars are close binaries, a recent new scenario suggests that the X-ray emission originates from the interaction between 
the wind from the companion star (a helium star or white dwarf) and the Be star disk and/or wind (see Langer et al. 2020). For $\gamma$~Cas, the estimated 
mechanical luminosity of the wind ($\sim$\,3.9\,$\times$\,10$^{34}$\,erg\,s$^{-1}$; see above) is much larger than its X-ray luminosity ($L_{\rm X}$) observed at 
the band of $\sim$\,0.5-10\,keV ($\sim$\,8.5\,$\times$\,10$^{32}$\,erg\,s$^{-1}$; or $\sim$\,2\% of the mechanical luminosity of the wind). 
Recent studies based on the X-ray surveys by the {\it XMM-Newton} satellite (covering 0.2-12\,keV) suggest a relation of 
\begin{equation}
{\rm log}~L_{\rm X}~=~23.6\pm3.8 + (0.3\pm0.1)~{\rm log}~L_{\rm wind} 
\end{equation}
for spectroscopically identified Galactic O stars (Nebot \& Oskinova 2018). 
For a comparison, an earlier survey by the {\it Einstein} satellite (covering 0.2-4.0\,keV) suggested another relation of
\begin{equation}
{\rm log}~L_{\rm X}~=~0.51^{+0.09} _{-0.12}~{\rm log}~L_{\rm wind} + 14.5^{+4.20}_{-3.30}
\end{equation}
for O stars (Sciortino et al. 1990).
Assuming that the relations also work for Be stars (note that {\bf the} parameter space in these relations is large), the estimated $L_{\rm X}$ ranges between roughly 
1.4\,$\times$\,10$^{32}$\,erg\,s$^{-1}$ ({\it Einstein} relation) and 9.5\,$\times$\,10$^{33}$\,erg\,s$^{-1}$ ({\it Newton} relation) for $\gamma$~Cas. The estimated 
values are comparable with those detected in the X-ray observations. This indicates that the mechanical luminosity of the wind $L_{\rm wind}$ is sufficient to account 
for the observed X-ray luminosity in $\gamma$~Cas (see also the discussion in Langer et al. 2020),  although the accurate conversion efficiency between 
$L_{\rm wind}$ and $L_{\rm X}$ is unclear.

In the $\gamma$~Cas binary system, the interaction between the primary and secondary stars was suggested by bow-shock nebula detected in the WISE 
22\,$\mu$m images (Bodensteiner et al. 2018; Langer et al. 2020). Therefore, the interaction between the winds from the primary and secondary stars can be 
expected in turn. 
Numerical studies have shown that the interaction between the winds in massive binary systems will produce much harder X-ray emission than that from 
single massive stars (Rauw \& Naz{\'e} 2016; Pittard \& Dawson 2018). Furthermore, wind-wind and wind-disk interactions in binary systems will result in 
multiple thermal components detected in the X-ray observations (see Hamaguchi et al. 2016; Langer et al. 2020). The short variability of the X-ray luminosity 
(in the time scale of seconds or minutes) seen in the $\gamma$~Cas stars can be explained by the high intrinsic instability of the wind, while the phase-dependent 
variability (hours or days) may be due to changes in the stellar separation, wind absorption and/or stellar occultation (Rauw \& Naz{\'e} 2016; Pittard \& Dawson 2018).

\section{Summary} \label{sec:summary}

We present wide-field HI4PI HI and MWISP CO images toward $\gamma$~Cas, in order to study its feedback toward the interstellar environment. 
A large expanding cavity, with a dimension of $\sim$\,2\fdg0\,$\times$\,1\fdg4 (or 6.0\,pc\,$\times$\,4.2\,pc), is discovered toward $\gamma$~Cas 
in the HI images. The measured expansion velocity is roughly 5.0\,$\pm$\,0.5\,\vkm. The comparison between multi-wavelengths images indicates that the cavity is 
opened by strong stellar wind from $\gamma$~Cas, and the estimated mechanical luminosity of the wind is $\sim$\,(3.9\,$\pm$\,1.0)\,$\times$\,10$^{34}$\,erg\,s$^{-1}$.
The CO observations show no clear molecular cloud formed on the dense shell of the cavity. Furthermore, strong CO emission is detected from IC\,63 and 
IC\,59, the two cometary globules illuminated by $\gamma$~Cas. The systematic velocity gradients ($\sim$\,20-30\,\vkm\,pc$^{-1}$) are observed in the two 
globules, proving the fast acceleration under the pressure from the stellar radiation. 
We consider that stellar wind from $\gamma$~Cas has high potential to lead to its observed peculiar X-ray emission. Our result favors a recent new scenario 
that emphasizes the roles of stellar wind and binarity in the X-ray emission for the $\gamma$~Cas stars.

\acknowledgments

We thank an anonymous referee for providing insightful suggestions and comments, which help us to improve this work.
This work was supported by the National Key R\&D Program of China (grant No. 2017YFA0402702), the National Natural Science Foundation of China 
(grant Nos. 12041305 and 11629302), and CAS International Cooperation Program (grant No. 114332KYSB20190009). 
This research is based on data products from the HI4PI, full-sky H$\alpha$ map, WISE, and MWISP surveys. We are grateful to all the members of the 
MWIPS CO line survey group, especially the staff of Qinghai Radio Station of PMO at Delingha for their support during the CO line observations.
MWISP is sponsored by National Key R\&D Program of China with grant 2017YFA0402701 and CAS Key Research Program of Frontier Sciences with 
grant QYZDJ-SSW-SLH047.

\clearpage

\appendix

\section{The HI4PI HI and MWISP CO maps of  $\gamma$~Cas}

Figure~5 shows the HI4PI line channel map toward $\gamma$~Cas at the velocity range from --20 to +5\,\vkm, where detailed HI gas structure and 
kinematics toward $\gamma$~Cas can be seen.
Figure~6 shows the MWISP CO line emission within the velocity range from --17 to +7\,\vkm, presenting the details of molecular gas surrounding
$\gamma$~Cas. Figure~7 shows the CO enlarged views toward the IC\,63 and IC\,59 globules.

\section{The Str{\"o}mgren sphere produced by $\gamma$~Cas}

After the formation of massive stars in the molecular cloud, the massive stars will form bright  HII regions through photoionization, which is firstly
introduced in detail by Str{\"o}mgren (1939). If a source of $Q_{\rm H}$ ionizing photons per second ignites in a cloud with an initial number density 
of $n_{\rm 0}$ atoms cm$^{-3}$, the initial number density of ions $n_{i}$ will be equal to $n_{\rm 0}$. If the cloud is pure hydrogen and completely 
electrically neutral, then the number density of electrons  $n_{\rm e}$ will be equal to the number density of ions, that is, 
we have $n_{\rm e}$  = $n_{i}$ = $n_{\rm 0}$.

When the ionization source is formed, the ions will recapture electrons in the ionization region and recombine to the bound state of the atom, which 
will consume part of the photons. Finally, the photons consumed by recombination and those produced by ionization source are balanced, and no 
more central gas is ionized. In this state, the Str{\"o}mgren sphere in the ionization region is formed, and the radius of the Str{\"o}mgren sphere can 
be expressed as
\begin{equation}\label{f.sphere}
 R_{\rm S}~ = ~\left ( \frac{3Q_{\rm H}}{4\pi \alpha_{\rm B} n_0^2} \right )^\frac{1}{3},
\end{equation}
where $\alpha_{\rm B}$\,=\,2.6\,$\times$\,10$^{-13}$\,(10$^4$\,K/$T$)$^{0.7}$\,cm$^{3}$\,s$^{-1}$ is the recombination rate at the ionized gas temperature 
of $T$. For $\gamma$~Cas (a bright B0.5\,IVe star), we take 7.9\,$\times$\,10$^{47}$ photons\,s$^{-1}$ as the stellar ionizing photon rate (see Vacca et al. 
1996), while the temperature of the ionized gas is assumed to be 10000\,K. The number density is estimated to be $\sim$\,150\,cm$^{-3}$ based on the HI 
observations (see $\S$\,3.1). Then, the estimated radius of the Str{\"o}mgren sphere $R_{\rm S}$ is roughly 1.0\,pc or $\sim$\,0\fdg33 at the distance of 
168\,pc.

It is also known that the Str{\"o}mgren sphere is expanding during the evolution of the ionization source. This process was first studied by Spitzer (1978), who 
derived the well-known expansion relation
\begin{equation}\label{f.spitzer}
 R(t) ~ = ~ R_{\rm S} \left ( 1 + \frac{7}{4} \frac{c_{\rm HII} t}{R_{\rm S}}\right )^\frac{4}{7},
\end{equation}
where $c_{\rm HII}$ is the sound speed in the ionized gas and $t$ is the evolution time of the HII region. 
In the calculations, the sound speed $c_{\rm HII}$ is generally assumed to be 10\,\vkm. The evolution time of the HII region is still uncertain for $\gamma$~Cas. 
Assuming $t$\,=\,1\,Myr (about 3 times of the dynamical age of the HI cavity, i.e., $\sim$\,3.0\,$\times$\,10$^5$\,yr; see $\S$\,3.1), the radius of the evolved
Str{\"o}mgren sphere is $\sim$\,5.3\,pc (or $\sim$\,1\fdg8), which is consistent with the result seen in the H$\alpha$ images (see Figure~1b).

\begin{figure}[ht!]
\plotone{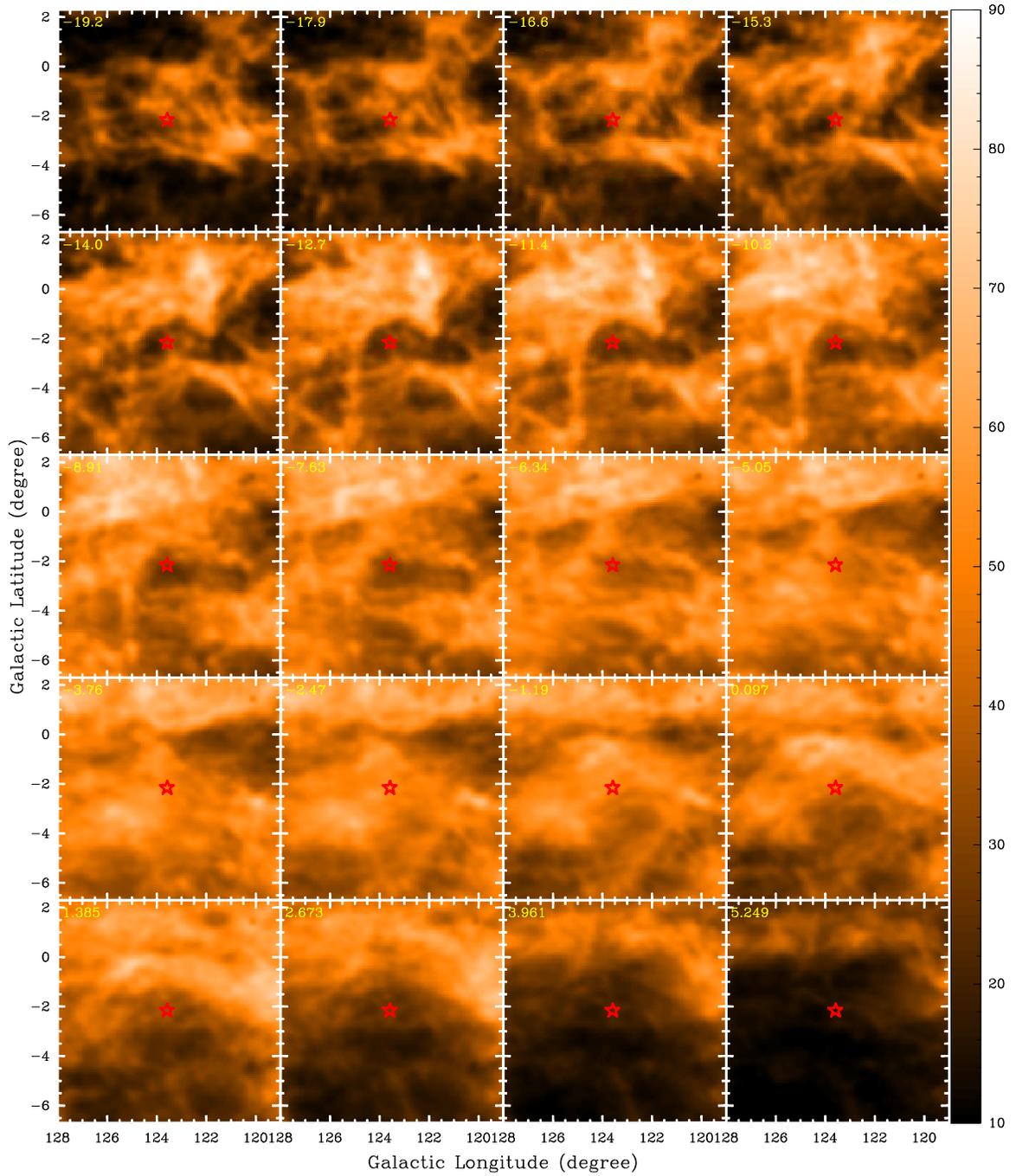}
\caption{The HI4PI HI line velocity channel maps toward $\gamma$~Cas. The unit of the scale bar is K. For all panels, red stellar symbol marks the 
position of $\gamma$~Cas. The LSR velocity is written in the top left corner of each panel (in km\,s$^{-1}$), and the systemic velocity of $\gamma$~Cas 
is approximately --10\,\vkm.\label{channel_HI}}
\end{figure}

\begin{figure}
\plotone{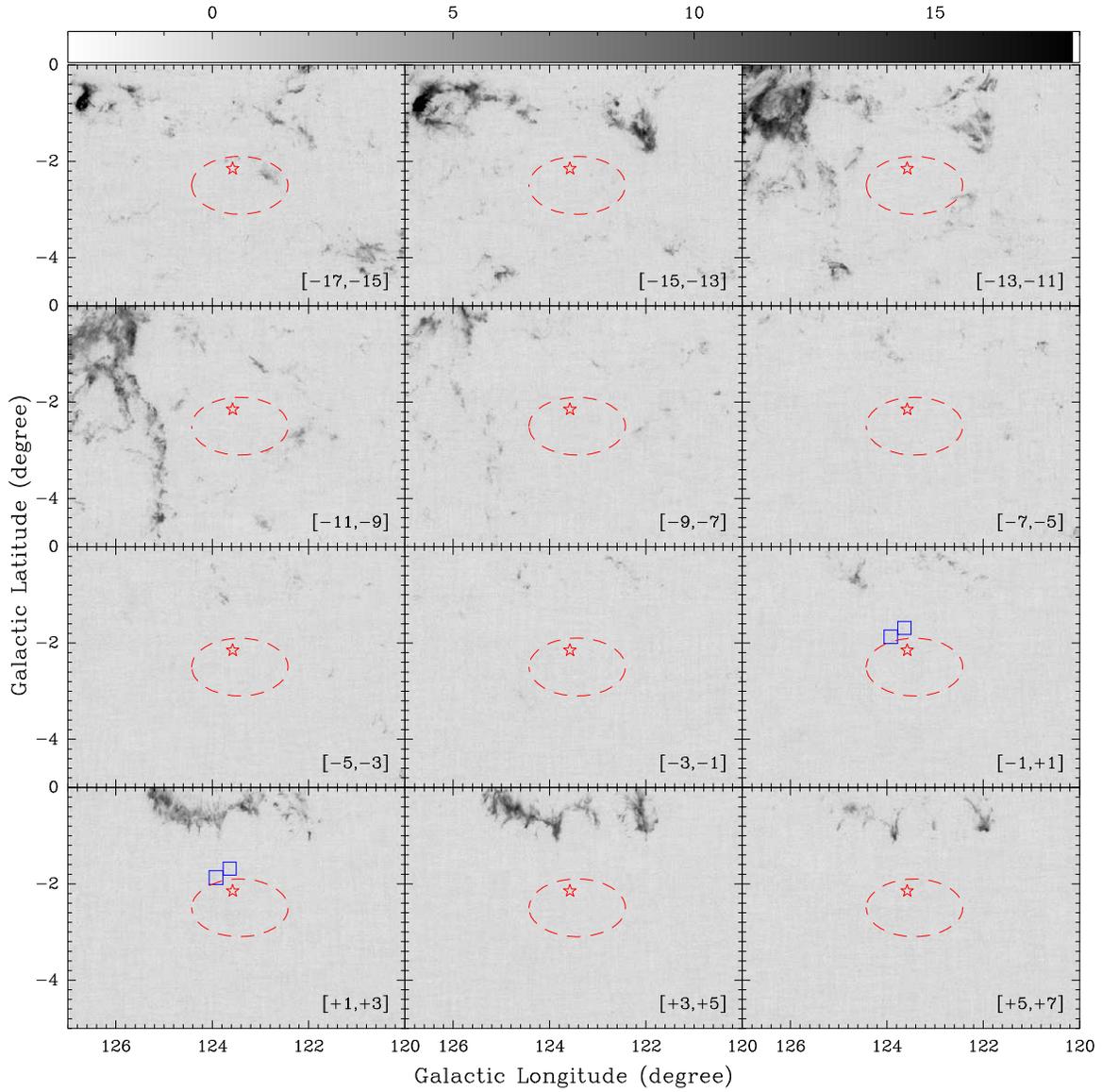}
\caption{The MWISP $^{12}$CO line velocity channel maps toward $\gamma$~Cas. The unit of the scale bar is K\,\vkm. For all panels, red stellar symbol 
marks the position of $\gamma$~Cas. The integrated velocity range is written in the bottom right corner of each panel (in km\,s$^{-1}$). The red dashed 
ellipse shows the cavity found in the HI observations. The blue squares show the extents of the IC\,63 and IC\,59 globules (see below Figure~7). 
\label{channel_CO}}
\end{figure}

\begin{figure}
\plotone{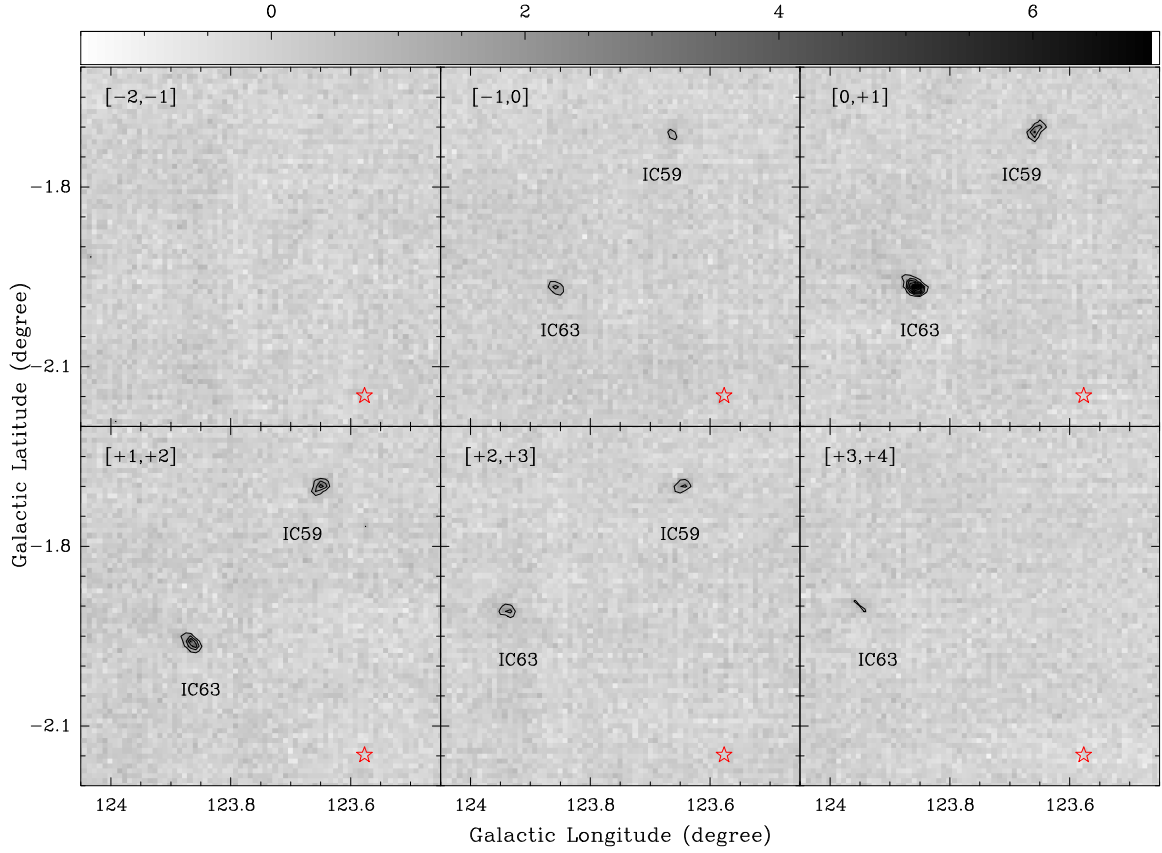}
\caption{The enlarged MWISP $^{12}$CO line velocity channel maps toward IC\,63 and IC\,59. The unit of the scale bar is K\,\vkm. For all panels, red 
stellar symbol marks the position of $\gamma$~Cas. The integrated velocity range is written in the top left corner of each panel (in km\,s$^{-1}$). The 
CO contours start from 1.0\,K\,\vkm\ (1\,$\sigma$\,$\sim$\,0.25\,K\,\vkm) and then increase by steps of 1.0\,K\,\vkm.\label{channel_globules}}
\end{figure}

\end{document}